\documentstyle[12pt]{article}
\begin{document}

\title{\Large\bf Spontaneous compactification of D=5 two-form gauge
fields and the obtainment of Maxwell and Yang-Mills theories}

\author{\\
J. Barcelos-Neto\thanks{\noindent
e-mail: barcelos@if.ufrj.br}\\
Instituto de F\'{\i}sica\\
Universidade Federal do Rio de Janeiro\\
RJ 21945-970 - Caixa Postal 68528 - Brasil\\}
\date{}

\maketitle
\abstract
We show that the spontaneous compactification of the Abelian and
non-Abelian two-form gauge field theories from $D=4+1$ to $D=3+1$
leads to the same theories plus the Maxwell and Yang-Mills ones,
respectively. The vector potential comes from the zero mode of the
fifth component of the tensor gauge field in $D=5$. Concerning to the
non-Abelian case, it is necessary to make a more refined definition
of the three-form stress tensor in order to be compatible, after the
compactification, with the two-form stress tensor of the Yang-Mills
theory.

\vspace{1cm}

\vspace{1cm}
\newpage

\section*{I. Introduction}
\renewcommand{\theequation}{I.\arabic{equation}}
\setcounter{equation}{0}

\bigskip
A significant number of quantum field theories we have to describe the
real world in $D=4$ are effective theories, in a sense that they
result from the absorption of some degrees of freedom of more general
theories. For example, the vector particles related to the weak force
are massful even though the corresponding gauge theory consider them
initially massless. This occurs because spontaneous symmetry breaking
together with Higgs mechanism leads to an effective theory where gauge
particles become actually massive.

\medskip
Another interesting aspect of mass generation for gauge fields, as it
was initially pointed out by Cremmer and Scherk \cite{Cremmer}, is by
means of a vector-tensor gauge theory \cite{Kalb} where these fields
are coupled in a topological way. Let me present some details of this
mechanism in order to make some comparison with the work we are going
to develop in this paper. The Lagrangian for the vector-tensor gauge
theory with topological coupling is given by (we consider the Abelian
case first)

\begin{equation}
{\cal L}=\frac{1}{12}\,H_{\mu\nu\rho}H^{\mu\nu\rho}
-\frac{1}{4}\,F_{\mu\nu}F^{\mu\nu}
+\frac{1}{2}\,m\,\epsilon_{\mu\nu\rho\lambda}\,
A^\mu\partial^\nu B^{\rho\lambda}
\label{I.1}
\end{equation}

\bigskip\noindent
where the antisymmetric stress tensors $H_{\mu\nu\rho}$ and
$F_{\mu\nu}$ are defined in terms of (antisymmetric) tensor and
vector potentials $B_{\mu\nu}$ and $A_\mu$ as

\begin{eqnarray}
H_{\mu\nu\rho}&=&\partial_\mu B_{\nu\rho}
+\partial_\rho B_{\mu\nu}+\partial_\nu B_{\rho\mu}
\nonumber\\
F_{\mu\nu}&=&\partial_\mu A_\nu-\partial_\nu A_\mu
\label{I.2}
\end{eqnarray}

\bigskip\noindent
This theory is invariant under the gauge transformations

\begin{eqnarray}
\delta B_{\mu\nu}&=&\partial_\mu\xi_\nu
-\partial_\nu\xi_\mu
\label{I.3a}\\
\delta A_\mu&=&\partial_\mu\epsilon
\label{I.3b}
\end{eqnarray}

\bigskip
If one considers the path integral formalism and integrates out the
tensor fields $B_{\mu\nu}$, the resulting effective theory is massive
for the vector field \cite{Allen,Barc1}. This can be considered as an
alternative mechanism of mass generation without Higgs bosons. The
non-Abelian version of this theory requires some care, because the
reducibility condition is only achieved in the vanishing surface of
the Maxwell stress tensor \cite{Freedmann}, or by using a kind of
Stuckelberg field \cite{Lahiri,Hwang}.

\bigskip
Another example that reinforces this point of view can be found in the
string scenario. It is well-known that consistent string theories can
only be formulated in spacetime dimensions higher than four.
Consequently, the theories we have to describe the world in $D=4$
might be effective theories of those ones formulated in, say, $D=10$
or $D=11$, and conveniently compactified to $D=4$. Of course, it is
not an easy task to know what are those original theories in $D=10$ or
$D=11$. However, it might be an interesting subject to investigate how
the theories we have in $D=4$ can come from extended theories
Ðormulated in spacetimes with dimensions higher than four. In this
sense we mention Kaluza-Klein \cite{Kaluza} that is formulated in $D=
5$ as a pure Einstein theory and gives Einstein and Maxwell theories
in $D=4$ (the gauge symmetry of the Maxwell theory is originated from
the fifth spacetime coordinate transformation).

\bigskip
The purpose of the present paper is to follow a similar procedure as
the Kaluza-Klein, but starting from a pure two-form gauge field theory
in $D=5$, both Abelian and non-Abelian. We show that, after
spontaneous compactification, Maxwell and Yang-Mills theories
naturally emerge as the zero mode of infinite Fourier excitations.
However, contrarily to Kaluza-Klein, the gauge symmetry does not come
from a spacetime coordinate transformation, but from the fifth
component of the tensor gauge symmetry. Another interesting aspect of
this mechanism is that photon and color fields remain massless when
tensor fields are integrated out (only higher excitations become
massive). We also show that the topological coupling term of
expression (\ref{I.1}) does not come from the compactification of any
term formulated at $D=5$.

\medskip
This paper is organized as follows: In Sec. II we develop the
compactification of the Abelian case, where the electromagnetic
Maxwell theory is obtained. In the first part of In Sec. III we
present some details of the non-Abelian formulation for the two-form
gauge theory. Our goal in this section is to figure out the action we
are going to compactify from $D=5$ to $D=4$ in order to obtain the
Yang-Mills theory. We show that it is necessary to make an appropriate
definition of the three-form stress tensor different of that one we
usually find in literature. Sec. IV contains the compactification of
the non-Abelian case and we left Sec. V for some concluding remarks.
We also include an Appendix to illustrate the Abelian compactification
in the language of differential forms.

\vspace{1cm}
\section*{II. Spontaneous compactification - Abelian case}
\renewcommand{\theequation}{II.\arabic{equation}}
\setcounter{equation}{0}

\bigskip
Let us start from the Lagrangian

\begin{equation}
{\cal L}=\frac{1}{12}\,H_{MNP}\,H^{MNP}
\hspace{1cm}M,N,P=0,\dots,4
\label{II.1}
\end{equation}

\bigskip\noindent
where we use capital indices to characterize the $D=5$ spacetime
components. The stress tensor $H_{MNP}$ is defined as in the first
relation (\ref{I.2}) and, consequently, the gauge transformation of
$B_{MN}$ is similar to the one given by (\ref{I.3a}).

\medskip
In order to perform the spontaneous compactification to $D=4$, which
is achieved by integrating out the coordinate $x_4$ in a circle of
radius $R$, we consider the tensor potential $B_{MN}$ split as

\begin{equation}
B_{MN}=(B_{\mu\nu},\,B_{4\mu})
\hspace{1cm}\mu,\nu=0,\dots,3
\label{II.2}
\end{equation}

\bigskip\noindent
So, we get the action

\begin{equation}
S=\int d^4x\int_0^Rdx^4\,\Bigl(\frac{1}{12}\,
H_{\mu\nu\rho}H^{\mu\nu\rho}
+\frac{1}{4}\,H_{4\mu\nu}H^{4\mu\nu}\Bigr)
\label{II.3}
\end{equation}

\bigskip\noindent
Developing the stress component $H_{4\mu\nu}$, we may write

\begin{eqnarray}
H_{4\mu\nu}&=&\partial_4B_{\mu\nu}
+\partial_\nu B_{4\mu}+\partial_\mu B_{\nu4}
\nonumber\\
&=&\partial_4B_{\mu\nu}
+\partial_\mu B_{\nu4}-\partial_\nu B_{\mu4}
\nonumber\\
&=&\partial_4B_{\mu\nu}+F_{\mu\nu}
\label{II.4}
\end{eqnarray}

\bigskip\noindent
where we have defined

\begin{equation}
B_{\mu4}=A_\mu
\label{II.5}
\end{equation}

\bigskip\noindent
with the purpose of making future comparisons in the Maxwell theory.
However, the quantity $A_\mu$ given by (\ref{II.5}) is not the photon
field yet. We notice that all fields in the action (\ref{II.3}) depend
on $x_\mu$ and $x_4$ and the gauge transformation of $A_\mu$ is not
the usual gauge transformation of the photon field (from (\ref{I.3b}),
we have $\delta A_\mu=\partial_\mu\xi_4-\partial_4\xi_\mu$).

\medskip
Using the result given by (\ref{II.4}), the general form of the action
turns to be

\begin{eqnarray}
S=\int d^4x\int_0^Rdx^4\,
\Bigl(\frac{1}{12}\,H_{\mu\nu\rho}H^{\mu\nu\rho}
-\frac{1}{4}\,F_{\mu\nu}F^{\mu\nu}
&\!\!\!+\!\!\!&\frac{1}{4}\,\partial_4B_{\mu\nu}\partial^4B^{\mu\nu}
\nonumber\\
&\!\!\!+\!\!\!&\frac{1}{2}\,F_{\mu\nu}\partial^4B^{\mu\nu}\Bigr)
\label{II.7}
\end{eqnarray}

\bigskip
The next step is to expand the fields $B_{\mu\nu}$ and $A_\mu$, as
well as the gauge parameters $\xi_\mu$ and $\xi_4$, in terms of
Fourier harmonics

\begin{eqnarray}
B_{\mu\nu}(x,x_4)&=&\frac{1}{\sqrt R}\sum_{n=-\infty}^{+\infty}
B_{(n)\,\mu\nu}(x)\,\exp\Bigl(2in\pi\frac{x^4}{R}\Bigr)\,,
\nonumber\\
A_\mu(x,x_4)&=&\frac{1}{\sqrt R}\sum_{n=-\infty}^{+\infty}
A_{(n)\,\mu}(x)\,\exp\Bigl(2in\pi\frac{x^4}{R}\Bigr)
\nonumber\\
\xi_\mu(x,x_4)&=&\frac{1}{\sqrt R}\sum_{n=-\infty}^{+\infty}
\xi_{(n)\,\mu}(x)\,\exp\Bigl(2in\pi\frac{x^4}{R}\Bigr)
\nonumber\\
\xi_4(x,x_4)&=&\frac{1}{\sqrt R}\sum_{n=-\infty}^{+\infty}
\xi_{(n)}(x)\,\exp\Bigl(2in\pi\frac{x^4}{R}\Bigr)
\label{II.8}
\end{eqnarray}

\bigskip\noindent
Since the fields $B_{\mu\nu}$ and $A_\mu$ are real, as well as the
gauge parameters, the Fourier modes must satisfy the conditions
$B^\dagger_{(n)\mu\nu}=B^\dagger_{(-n)\mu\nu}$, $A^\dagger_{(n)\mu}=
A_{(-n)\mu}$, etc., where dagger means complex conjugation.

\medskip
Developing the terms of the action (\ref{II.7}) by using the
expansions given by (\ref{II.8}), we obtain

\begin{eqnarray}
S=\int d^4x\sum_{n=0}^\infty\Bigl(
\frac{1}{12}\,H_{(n)\,\mu\nu\rho}H_{(n)}^{\dagger\,\mu\nu\rho}
-\frac{1}{4}\,F_{(n)\,\mu\nu}F_{(n)}^{\dagger\,\mu\nu}
&\!\!\!+\!\!\!&\frac{\pi^2n^2}{R^2}\,
B_{(n)\,\mu\nu}B_{(n)}^{\dagger\,\mu\nu}
\nonumber\\
&\!\!\!-\!\!\!&\frac{i\pi n}{R}\,
F_{(n)\,\mu\nu}B_{(n)}^{\dagger\,\mu\nu}\Bigr)
\label{II.9}
\end{eqnarray}

\bigskip\noindent
which is invariant for the gauge transformations

\begin{eqnarray}
\delta B_{(n)\,\mu\nu}&=&\partial_\mu\xi_{(n)\,\nu}
-\partial_\nu\xi_{(n)\,\mu}
\label{II.10}\\
\delta A_{(n)\,\mu}&=&\partial_\mu\xi_{(n)}
-\frac{2i\pi n}{R}\,\xi_{(n)\,\mu}
\label{II.11}
\end{eqnarray}

\bigskip
We then notice that $-\frac{1}{4}\,F_{(0)\,\mu\nu}F_{(0)}^{\mu\nu}$ is
the Maxwell Lagrangian. In fact, from (\ref{II.11}) we have that the
gauge transformation of $A_{(0)\mu}$ is just $\partial_\mu\xi_{(0)}$.
Further, contrarily to the vector-tensor gauge theory in $D=4$, the
photon field $A_{(0)\mu}$ does not acquire mass after integrating over
the tensor field $B_{(0)\mu\nu}$ (for $n=0$, $A_{(n)\mu}$ and $B_{(n)
\mu\nu}$ decouple). However, higher excitations are massive.

\vspace{1cm}
\section*{III. Non-Abelian formulation of the two-form gauge theory}
\renewcommand{\theequation}{III.\arabic{equation}}
\setcounter{equation}{0}

\bigskip
We start this section by reviewing the main aspects of the non-Abelian
two-form gauge field theory. We shall see that there is an
arbitrariness in defining the corresponding field strength (what does
not happen in the Abelian counterpart). The definition that usually
appears in literature \cite{Lahiri,Hwang} is not in agreement, after
the compactification, with the Yang-Mills theory. We find this an
important point because the coherency between compactification and the
correct obtainment of the Yang-Mills theory might be the guidance to a
precise definition of the non-Abelian field strength tensor.

\medskip
From this section on, we opt to work with differential forms because
the notation is simpler and it is easier to make comparations between
one and two-form gauge theories. We found convenient does not work
with differential forms in the previous sections because this compact
notation would hidden some details we would like to emphasize at that
opportunity. We display in the Appendix A the compactification of the
Abelian case in the language of differential forms.

\medskip
Let us first consider the one-form case. We then start from the
introduction of the one-form connection

\begin{eqnarray}
\Gamma&=&A_\mu\,dx^\mu
\nonumber\\
&=&A_\mu^aT^a\,dx^\mu
\label{III.1}
\end{eqnarray}

\bigskip\noindent
that is a Lie algebra valued on the $SU(N)$ symmetry group
($a=1,\dots,N^2-1$), whose generators satisfy

\begin{eqnarray}
[T^a,T^b]&=&i\,f^{abc}T^c
\nonumber\\
{\rm Tr}\,(T^aT^b)&=&\frac{1}{2}\delta^{ab}
\nonumber\\
(T^a)^{bc}&=&-\,if^{abc}
\label{III.2}
\end{eqnarray}

\bigskip\noindent
Of course, in the definition of a differential $p$-form, the spacetime
can have any dimension $D$ with $p\leq D$. In this brief review, we
take the usual spacetime dimension $D=4$ what is implicit in the use
of Greek indices.

\medskip
The connection permit us to define the exterior covariant derivative
as \cite{Eguchi}

\begin{eqnarray}
D\omega&=&d\omega-i\Gamma\wedge\omega
+i(-1)^p\omega\wedge\Gamma
\nonumber\\
&\equiv&d\omega-i[\Gamma,\omega]
\label{III.3}
\end{eqnarray}

\bigskip\noindent
where $\omega$ is a Lie algebra valued $p$-form
($\omega=\omega^aT^a$) and $d$ represents the usual exterior
derivative.

\medskip
The curvature two-form is defined to be

\begin{equation}
F=d\Gamma-i\,\Gamma\wedge\Gamma
\label{III.4}
\end{equation}

\bigskip\noindent
It is important to observe that $F$ is not the covariant derivative
of $\Gamma$. At this point resides the arbitrariness in the
definition of the three-form strength as we are going to see soon.

\medskip
The definition of the exterior derivative and the curvature two-form
permit us to introduce the Bianchi identities

\begin{eqnarray}
DD\,\omega&=& i\,[\omega,F]
\label{III.5}\\
DF&=&0
\label{III.6}
\end{eqnarray}

\bigskip\noindent
that are satisfied for any gauge connection $\Gamma$ and any algebra
valued $p$-form $\omega$. A fundamental consequence of (\ref{III.5})
is that if one defines the gauge variation of the one-form connection
like

\begin{equation}
\delta\Gamma=D\epsilon
\label{III.7}
\end{equation}

\bigskip\noindent
where $\epsilon$ is an infinitesimal Lie-algebra valued zero-form
parameter ($\epsilon=\epsilon^aT^a$), the curvature two-form
transforms as

\begin{eqnarray}
\delta F&=&d\delta\Gamma-i\delta\Gamma\wedge\Gamma
+i\Gamma\wedge\delta\Gamma
\nonumber\\
&=&D\delta\Gamma
\nonumber\\
&=&DD\epsilon
\nonumber\\
&=&i\,[\epsilon,F]
\label{III.8}
\end{eqnarray}

\bigskip\noindent
We observe, in the second step above, that $\delta F$ is the covariant
derivative of $\delta\Gamma$, even though $F$ does not have this
property with respect to $\Gamma$.

\medskip
The result (\ref{III.8}) implies that the action

\begin{equation}
S=-\frac{1}{2}\,{\rm Tr}\int F\wedge{^\ast F}
\label{III.9}
\end{equation}

\bigskip\noindent
is gauge invariant, due to the cyclic property of the trace operation.
In (\ref{III.9}), the symbol $^\ast$ represents the Hodge duality
operation. So, the integrand is proportional to the oriented volume
element in the Minkowiski space-time. To be more precise, the duality
operation maps the p-form coordinate basis $\{1,dx^\mu,dx^\mu\wedge
dx^\nu,dx^\mu\wedge dx^\nu\wedge dx^\rho,dx^\mu\wedge dx^\nu\wedge
dx^\rho\wedge dx^\sigma\}$ into the basis
$\{\eta,\eta^\mu,\eta^{\mu\nu},\eta^{\mu\nu\rho},
\eta^{\mu\nu\rho\sigma}\}$. In these expressions, $\eta$ is the four-form oriented volume
element, $\eta^\mu$ is a three-form, $\eta^{\mu\nu}$ is a two-form and
so on.  They satisfy relations such $dx^\mu\wedge\eta_\nu=
\delta^\mu_\nu\eta\,,dx^\mu\wedge\eta_{\nu\rho} =2\delta^\mu_{[\nu}
\eta_{\rho]}$ and $dx^\mu\wedge\eta_{\nu\rho\sigma} =
3\delta^\mu_{[\nu}\eta_{\rho\sigma]}$.  As $F={1\over2}F_{\mu\nu}
dx^\mu\wedge dx^\nu$, $^\ast F={1\over2}F_{\mu\nu}\eta^{\mu\nu}$ and
consequently $F\wedge^\ast F= {1\over2}F_{\mu\nu}F^{\mu\nu}\eta$.

\bigskip
Let us now consider the non-Abelian two-form case. We start by
introducing a two-form Lie-algebra valued object $\Lambda$ in a
similar way it was done for the connection $\Gamma$, i.e.,

\begin{equation}
\Lambda=\frac{1}{2}\,B^a_{\mu\nu}T^a\,dx^\mu\wedge dx^\nu
\label{III.10}
\end{equation}

\bigskip\noindent
Even though $\Gamma$
is not a connection, it looks like natural to assume its gauge
transformation as being

\begin{equation}
\delta_T\Lambda=D\xi
\label{III.11}
\end{equation}

\bigskip\noindent
where the subscript $T$ means that the transformation above is just
part (related to the tensor sector) of a more general transformation
as we are going to see just below. Here, $\xi$ is an infinitesimal
Lie-albegra valued one-form gauge parameter.

\medskip
We see that the gauge transformation (\ref{III.11}) is a natural
extension of (\ref{I.3a}) and (\ref{III.7}). However, contrarily to
the Abelian two-form case, it is not reducible. In fact, if one takes
the one-form parameter $\xi$ as the (covariant) derivative of a
zero-form parameter, say $\alpha$, we find

\begin{eqnarray}
\delta_T\Lambda&=&DD\alpha
\nonumber\\
&=&i\,[\alpha,F]
\label{III.12}
\end{eqnarray}

\bigskip\noindent
where in the last step we have used the Bianchi identity
(\ref{III.5}). We notice that the reducibility condition is only
attained if the curvature $F$ vanishes identically \cite{Freedmann}.

\bigskip
Since $\Lambda$ is a Lie-algebra valued object, it may couple with the
connection $\Gamma$ and, consequently, it can have an additional
transformation related to the vector sector. One considers that this
additional transformation is given by (see expression \ref{III.8})

\begin{equation}
\delta_V\Lambda=i\,[\epsilon,\Lambda]
\label{III.13}
\end{equation}

\bigskip\noindent
So, the general gauge transformation for $\Lambda$ is

\begin{equation}
\delta\Lambda=i\,[\epsilon,\Lambda]+D\xi
\label{III.14}
\end{equation}

\bigskip
Now, a controversial point is to define the object that will be the
extension of $F$. In the Abelian case, this is very simple and direct
because since $F$ is just the exterior derivative of $\Gamma$, it is
natural to assume that the extension of $F$, that we call $H$, is the
exterior derivative of $\Lambda$. However, in the non-Abelian case,
$F$ is not the covariant derivative of $\Gamma$. Hence, it is not
clear what should be $H$ in this case. What is usually done in
literature is to define this stress tensor as the covariant derivative
of $\Lambda$, even though $F$ does not have this property with respect
to $\Gamma$. Let us then see what happens if this definition is taken,
i.e.

\begin{equation}
H=D\Lambda
\label{III.15}
\end{equation}

\bigskip\noindent
Using (\ref{III.7}) and (\ref{III.14}), we obtain that the gauge
transformation for $H$ reads

\begin{equation}
H=i\,[\xi,F]+i\,[\epsilon,H]
\label{III.16}
\end{equation}

\bigskip\noindent
We notice that an action for $H$ similar to (\ref{III.9}), i.e.
$-\frac{1}{2}\,{\rm Tr}\int H\wedge{^\ast H}$, will be invariant for
the second part of (\ref{III.16}), but not for the first.

\medskip
This initial problem can be circumvented by redefining the two-form
$\Lambda$ as \cite{Lahiri,Hwang}

\begin{equation}
\tilde\Lambda=\Lambda+D\Omega
\label{III.17}
\end{equation}

\bigskip\noindent
where the one-form quantity $\Omega$ plays the role of a Stuckelberg
field. Considering that $\Omega$ has the gauge transformation

\begin{equation}
\delta\Omega=i\,[\epsilon,\Omega]-\xi
\label{III.18}
\end{equation}

\bigskip\noindent
we obtain that the gauge transformation for $\tilde\Lambda$ reads

\begin{equation}
\delta\tilde\Lambda=i\,[\epsilon,\tilde\Lambda]
\label{III.19}
\end{equation}

\bigskip\noindent
Keeping the definition that $\tilde H$ is the covariant derivative of
$\tilde\Lambda$, we have

\begin{equation}
\delta\tilde H=i\,[\epsilon,\tilde H]
\label{III.20}
\end{equation}

\bigskip\noindent
Now, an action like $-\frac{1}{2}\,{\rm Tr}\int\tilde
H\wedge{^\ast\tilde H}$ is gauge invariant.

\bigskip
The problem in defining $H$ as the covariant derivative of $\Lambda$
(or $\tilde H$ in terms of $\tilde\Lambda$) is that the Yang-Mills
theory is not obtained after the compactification from $D=5$ to $D=4$.
This is so because $F$ is not attained from $H$ after the
compactification. In fact, in the definition of $F$ we have the
product $\Gamma\wedge\Gamma$, while in the case of $H$ (or $\tilde H$)
the corresponding product is $[\Gamma,\Lambda]$ (or
$[\Gamma,\tilde\Lambda]$) (there is a factor 2 that spoils the correct
obtainment of $F$).

\medskip
This problem can also be circumvented by introducing another
Stuckelberg like field in the definition of the three-form stress
tensor. Denoting this quantity by $\tilde{\tilde H}$, and considering
the definition

\begin{equation}
\tilde{\tilde H}=d\tilde\Lambda
-\frac{i}{2}\,\Bigl[\Lambda+\Xi,\tilde\Lambda\Bigr]
\label{III.21}
\end{equation}

\bigskip\noindent
we have that an action like

\begin{equation}
S=-\frac{1}{2}\,{\rm Tr}\int\tilde{\tilde H}
\wedge{^\ast\tilde{\tilde H}}
\label{III.22}
\end{equation}

\bigskip\noindent
will be gauge invariant if $\Xi$ transforms as

\begin{equation}
\delta\Xi=d\epsilon-i\,[\Xi,\epsilon]
\label{III.23}
\end{equation}

\bigskip\noindent
This will be the action we are going to use in the compactification
from $D=5$ to $D=4$ in order to obtain the Yang-Mills theory.

\medskip
It might be opportune to mention that the use of two auxiliary
Stuckelberg fields is not new. They have already been introduced with
the purpose of restoring the reducible condition in the non-Abelian
sector \cite{Lahiri}, that is a different purpose of the use we are
making here.

\vspace{1cm}
\section*{IV. Spontaneous compactification of the Non-Abelian case}
\renewcommand{\theequation}{IV.\arabic{equation}}
\setcounter{equation}{0}

\bigskip
For comparison, see a similar development of the Abelian case in the
Appendix A. We start from the action

\begin{equation}
S=-\,\frac{1}{2}\,{\rm Tr}\int_{M_5}\tilde{\tilde{\bf H}}
\wedge{^\ast\tilde{\tilde{\bf H}}}
\label{IV.1}
\end{equation}

\bigskip\noindent
where

\begin{eqnarray}
\tilde{\tilde{\bf H}}&=&{\bf d}\tilde{\bf\Lambda}
-\frac{i}{2}\,[{\bf\Gamma}+{\bf\Xi},\tilde{\bf\Lambda}]
\label{IV.2}\\
\tilde{\bf\Lambda}&=&{\bf\Lambda}+{\bf D}{\bf\Omega}
\label{IV.3}
\end{eqnarray}

\bigskip\noindent
We are using the boldface notation to represent the geometrical
elements in $M_5$. Following the same steps of the Abelian case, we
isolate the $dx^4$ component from the quantities above. First, we take
${\bf\Lambda}$, ${\bf\Omega}$, ${\bf\Gamma}$, and ${\bf\Xi}$, and
introduce some definitions for the $dx^4$ component,

\begin{eqnarray}
{\bf\Lambda}&=&\Gamma\wedge dx^4+\Lambda
\nonumber\\
{\bf\Omega}&=&\varphi\,dx^4+\Omega
\nonumber\\
{\bf\Gamma}&=&\phi\,dx^4+\Gamma
\nonumber\\
{\bf\Xi}&=&\chi\,dx^4+\Xi
\label{IV.4}
\end{eqnarray}

\bigskip\noindent
where $\Lambda$, $\Gamma$, $\varphi$, $\Omega$, $\phi$, $\chi$, and
$\Xi$ depend on $(x^\mu,x^4)$. Consequently, we have

\begin{eqnarray}
{\bf D\Omega}&=&\Bigl(-\,\partial_4\Omega
+i\,[\phi,\Omega]+D\varphi\Bigr)\wedge dx^4+D\Omega
\nonumber\\
\tilde{\bf\Lambda}&=&\Bigl(\Gamma-\partial_4\Omega
+i\,[\phi,\Omega]+D\varphi\Bigr)\wedge dx^4
+\tilde\Lambda
\nonumber\\
{\bf d}\tilde{\bf\Lambda}&=&\Bigl(\partial_4\tilde\Lambda
+d\Gamma-d\partial_4\Omega+i\,[d\phi,\Omega]
+i\,[\phi,d\Omega]+dD\varphi\Bigr)\wedge dx^4
+d\tilde\Lambda
\label{IV.5}
\end{eqnarray}

\bigskip\noindent
The combination of (\ref{IV.4}) and (\ref{IV.5}) gives

\begin{equation}
\tilde{\tilde{\bf H}}=\tilde H+F\wedge dx^4+G\wedge dx^4
\label{IV.6}
\end{equation}

\bigskip\noindent
where $G$ is a compact notation for

\begin{eqnarray}
G&=&\partial_4\tilde\Lambda
+d\Bigl(D\varphi-\partial_4\Omega+i\,[\phi,\Omega]\Bigr)
-\frac{i}{2}\,[\phi+\chi,\tilde\Lambda]
\nonumber\\
&&\phantom{\partial_4\tilde\Lambda}
-\frac{i}{2}\,[\Gamma+\Xi,D\varphi-\partial_4\Omega
+i\,[\phi,\Omega]
\label{IV.7}
\end{eqnarray}

\bigskip\noindent
Now, the first Fourier component of $F$, that appears in (\ref{IV.6}),
can be identified as the Yang-Mills stress tensor. It is important to
emphasize that this was actually possible by virtue of the factor
$\frac{1}{2}$ we have introduced in the definition of
$\tilde{\tilde{\bf H}}$.

\medskip
The Hodge duality $^\ast{\tilde{\tilde{\bf H}}}$ is given by

\begin{equation}
^\ast{\tilde{\tilde{\bf H}}}=-\,^\ast{\tilde{\tilde H}}
\wedge dx^4+{^\ast F}+{^\ast G}
\label{IV.8}
\end{equation}

\medskip
Developing the quantities above in terms of Fourier harmonics and
replace them in the action (\ref{IV.1}), we easily obtain the Yang-
Mills theory from the first harmonic component when the coordinate
$x_4$ is integrated out.

\vspace{1cm}
\section*{V. Conclusion}

\bigskip
In this paper, we have studied the spontaneous compactification of the
two-form gauge field theory from $D=5$ to $D=4$. In the Abelian case,
this leads to the same theory plus Maxwell one. However, in the non-
Abelian case, the Yang-Mills theory is only attained if we make a
convenient new definition of the non-Abelian three-form stress tensor.

\medskip
To conclude, let us say that the topological term which couples vector
and tensor gauge fields in $D=4$, given at Eq. (\ref{I.1}), cannot be
generated from compactification. At first sight, we could think that
it is originated from a Chern-Simon term in $D=5$ like
$\kappa\,\epsilon^{MNPQR}\,\partial_MB_{NP}B_{QR}$. But we can
directly verify that this term is zero. We may then conclude that the
topological term of Eq. (\ref{I.1}) has its own origin just in $D=4$.
In a physical point of view, there are two explanations for this fact.
First, we know that the topological term in $D=4$ is the starting
point to generate mass for the vector potential if tensor degrees of
freedom are integrated out. On the other hand, if one starts from the
pure tensor gauge theory in $D=5$ and integrated out the $x_4$
component, the excitations for $n>0$ are already massive, without
necessity of any topological coupling terms. Another possibility is
that this term may have a quantum origin like the usual Chern-Simon
term in $D=3$ \cite{Redclich}. This second possibility is presently
under study, and possible results shall be reported elsewhere
\cite{Barc2}.

\vspace{1cm}
\noindent
{\bf Acknowledgment:} This work is supported in part by Conselho
Nacional de Desenvolvimento Cient\'{\i}fico e Tecnol\'ogico - CNPq,
Financiadora de Estudos e Projetos - FINEP and Funda\c{c}\~ao
Universit\'aria Jos\'e Bonif\'acio - FUJB (Brazilian Research
Agencies). I am in debt with R. Amorim for many useful discussions
during the elaboration of this work.

\vspace{1cm}
\appendix
\renewcommand{\theequation}{A.\arabic{equation}}
\setcounter{equation}{0}
\section*{Appendix A}

\bigskip
In this Appendix we consider the spontaneous compactification of the
Abelian case in the language of differential forms. First we
introduce the quantity

\begin{equation}
{\bf\Lambda}=\frac{1}{2}\,B_{MN}\,dx^M\wedge dx^N
\label{A.1}
\end{equation}

\bigskip\noindent
Let us rewrite it by isolating the $dx^4$ component (that shall be
integrated out in a circle)

\begin{eqnarray}
{\bf\Lambda}&=&B_{\mu4}(x,x_4)\,dx^\mu\wedge dx^4
+\frac{1}{2}\,B_{\mu\nu}(x,x_4)\,dx^\mu\wedge dx^\nu
\nonumber\\
&=&A_\mu(x,x_4)\,dx^\mu\wedge dx^4
+\frac{1}{2}\,B_{\mu\nu}(x,x_4)\,dx^\mu\wedge dx^\nu
\nonumber\\
&=&\Gamma(x,x_4)\wedge dx^4+\Lambda(x,x_4)
\label{A.2}
\end{eqnarray}

\bigskip\noindent
where $\Gamma$ and $\Lambda$ are differential forms in $M_4$, but they
do not correspond to one and two-forms of the Maxwell and tensor gauge
theories, respectively, because they still depend on the
coordinate $x_4$.
\medskip
Using the expression (\ref{A.2}), we calculate {\bf H} by means of the
following relation

\begin{eqnarray}
{\bf H}&=&{\bf d\Lambda}
\nonumber\\
&=&\bigl(dx^4\partial_4+dx^\mu\partial_\mu\bigr)
\wedge\bigl(\Gamma\wedge dx^4+\Lambda\bigr)
\nonumber\\
&=&dx^4\wedge\partial_4\Lambda
+dx^\mu\wedge\partial_\mu\Gamma\wedge dx^4
+dx^\mu\wedge\partial_\mu\Lambda
\nonumber\\
&=&\partial_4\Lambda\wedge dx^4+F\wedge dx^4+H
\label{A.3}
\end{eqnarray}

\bigskip\noindent
To construct the action, we need the dual $^\ast{\bf H}$, that
directly obtained by

\begin{eqnarray}
^\ast{\bf H}&=&^\ast\bigl({\bf d\Lambda}\bigr)
\nonumber\\
&=&\frac{1}{4}\,\epsilon^{MNPQ}\,\partial_MB_{NP}\,dx_Q\wedge dx_R
\nonumber\\
&=&\partial_4{^\ast\Lambda}+{^\ast F}-{^\ast H}\wedge dx^4
\label{A.4}
\end{eqnarray}

\bigskip\noindent
where $^\ast\Lambda$, $^\ast F$, and $^\ast H$, even though depend on
$x_4$, are Hodge dualities in $M_4$. Using the expressions for {\bf
H} and $^\ast${\bf H} given by the expressions above, we have the
action

\begin{eqnarray}
S&=&\frac{1}{2}\int_{M_5}{\bf H}\wedge^\ast{\bf H}
\nonumber\\
&=&\frac{1}{2}\int_{M_5}\Bigl(
\partial_4\Lambda\wedge\partial_4{^\ast\Lambda}
+2F\wedge\partial_4{^\ast\Lambda}
+F\wedge{^\ast F}-H\wedge{^\ast H}\Bigr)\wedge dx^4
\label{A.5}
\end{eqnarray}

\bigskip
The next step is to integrate the coordinate $x^4$ over a circle of
radius $R$. We then consider the following expansion of the forms
$\Lambda$, $F$, and $H$, as well as their Hodge dualities, in terms
of Fourier harmonics

\begin{eqnarray}
\Lambda(x,x_4)&=&\frac{1}{\sqrt R}\,
\sum_{n=-\infty}^{n=+\infty}\Lambda_{(n)}\,
\exp\Bigl(2in\pi\frac{x^4}{R}\Bigr)
\nonumber\\
^\ast\Lambda(x,x_4)&=&\frac{1}{\sqrt R}\,
\sum_{n=-\infty}^{n=+\infty}\,^\ast\Lambda_{(n)}\,
\exp\Bigl(2in\pi\frac{x^4}{R}\Bigr)
\nonumber\\
{\rm etc.}\label{A.6}
\end{eqnarray}

\bigskip\noindent
Introducing these expansions into the expression (\ref{A.5}) and
integrating out the coordinate $x^4$ on a circle of radius $R$, we
obtain

\begin{eqnarray}
&&S=\sum_{n=-\infty}^{n=+\infty}\int_{M_4}\biggl[
-\frac{1}{2}\,H_{(n)}\wedge{^\ast H_{(-n)}}
+\frac{1}{2}\,F_{(n)}\wedge{^\ast F_{(-n)}}
\nonumber\\
&&\phantom{S=\sum_{n=-\infty}^{n=+\infty}\int_{M_4}\biggl[}
\frac{1}{2}\Bigl(\frac{2n\pi}{R}\Bigr)^2\,
\Lambda_{(n)}\wedge{^\ast\Lambda_{(-n)}}
-\frac{2in\pi}{R}\,F_{(n)}\wedge{^\ast\Lambda_{(-n)}}\biggr]
\label{A.7}
\end{eqnarray}

\bigskip\noindent
Since $\Lambda$ and $^\ast\Lambda$ are real quantities, we have that
$\Lambda_{(-n)}=\Lambda^\dagger_{(n)}$ and
$^\ast\Lambda_{(-n)}=^\ast\Lambda^\dagger_{(n)}$.

\end{document}